*Article*

# Classification of Adventitious Sounds Combining Cochleogram and Vision Transformers

Loredana Daria Mang [1,*], Francisco David González Martínez [1], Damian Martinez Muñoz [1], Sebastián García Galán [1] and Raquel Cortina [2]

1 Department of Telecommunication Engineering, University of Jaen, 23700 Linares, Spain; fdgonzal@ujaen.es (F.D.G.M.); damian@ujaen.es (D.M.M.); sgalan@ujaen.es (S.G.G.)
2 Department of Computer Science, University of Oviedo, 33003 Oviedo, Spain; raquel@uniovi.es
* Correspondence: lmang@ujaen.es

**Abstract:** Early identification of respiratory irregularities is critical for improving lung health and reducing global mortality rates. The analysis of respiratory sounds plays a significant role in characterizing the respiratory system's condition and identifying abnormalities. The main contribution of this study is to investigate the performance when the input data, represented by cochleogram, is used to feed the Vision Transformer (ViT) architecture, since this input–classifier combination is the first time it has been applied to adventitious sound classification to our knowledge. Although ViT has shown promising results in audio classification tasks by applying self-attention to spectrogram patches, we extend this approach by applying the cochleogram, which captures specific spectro-temporal features of adventitious sounds. The proposed methodology is evaluated on the ICBHI dataset. We compare the classification performance of ViT with other state-of-the-art CNN approaches using spectrogram, Mel frequency cepstral coefficients, constant-Q transform, and cochleogram as input data. Our results confirm the superior classification performance combining cochleogram and ViT, highlighting the potential of ViT for reliable respiratory sound classification. This study contributes to the ongoing efforts in developing automatic intelligent techniques with the aim to significantly augment the speed and effectiveness of respiratory disease detection, thereby addressing a critical need in the medical field.

## 1. Introduction

Early detection of respiratory anomalies is key for taking care of the health status of the respiratory system, considering that the lungs are exposed to the external environment [1], making anyone who breathes vulnerable to becoming ill. The World Health Organization (WHO) has indicated that lung disorders remain one of the dominant causes of death rates around the world, being responsible for a significant impact on people's quality of life. The detection of these respiratory anomalies is of great importance to ensure that patients receive the right treatment as early as possible, which can significantly improve health outcomes and reduce the risk of complications. In addition, these types of anomalies, implicitly respiratory diseases, generate an immense global health burden, and most deaths occur in countries with scarce economic resources, so early diagnosis and rapid treatment remain critical to success in this area. In 2017, the top respiratory diseases were reported [1]: chronic obstructive pulmonary disease (COPD), asthma, acute respiratory infections (e.g., pneumonia), tuberculosis (TB), and lung cancer (LC). In this regard, COPD was responsible for more than 3 million deaths in 2019, of which the majority occurred in people under the age of 70 living in countries with emerging economies, as mentioned above [2]. Although asthma affected more than 250 million people in 2019 and caused more than 400,000 deaths, most of people who suffer from asthma do not have access to effective medications, especially in resource-limited countries [3]. Pneumonia was the cause of nearly 1 million deaths in children in 2017, highlighting adults over 65 and people with persistent pulmonary disorders as the group at risk [4]. TB killed more than 1.5 million people in 2021, with nearly USD 13 billion invested in its detection and treatment to reach the target proposed at the UN high level-meeting in 2018 [5]. LC is one of the deadliest cancers, killing 2.21 million people in 2020 [6]. As the health alarm associated with this type of diseases is becoming increasingly worrisome at the global level, researchers in artificial intelligence (AI) and signal processing are focused on developing automatic respiratory signal analysis systems, as it is a current challenge to apply these techniques to facilitate physicians' early detection of respiratory abnormalities and provide the patient with the right treatment at the right time, but without replacing the medical professional's diagnosis in healthcare. However, it is becoming increasingly clear that the use of these techniques allows for more effective signal analysis, enabling physicians to identify patterns and trends that may not be obvious to the naked eye and may be revealing the presence of a pulmonary disorder.

Although in the field of signal processing, deep learning techniques have been used in many fields, such as speech signals [7], image inpainting [8] or the diagnosis of Alzheimer's disease [9], the characterization of respiratory sounds is considered a relevant stage in order to model and extract clinically relevant information regarding the respiratory system's condition. Lung sounds can be categorized into two groups, normal lung sounds (RS) and abnormal or adventitious sounds (AS) [10], since the presence of AS usually suggests the existence of inflammation, infection, blockage, narrowing, or fluid in the lungs. In particular, RS are indicative of undamaged respiratory physiology and are usually heard in healthy lungs. These sounds show a broadband distribution in frequency, with the predominant energy located in the spectral band [100–2000] Hz [11]. Conversely, AS are observed in a time-frequency (TF) overlapping with RS and are typically present in people with any lung disease, including wheezes and crackles, which are more perceptible AS. Wheeze sounds (WS) are a type of continuous and tonal sound biomarker, displaying narrowband spectral trajectories with the fundamental frequency ranges from 100 Hz to 1000 Hz and at least a minimum duration of 80 ms [12,13] or 100 ms [10,13], generally associated with COPD and Asthma [14]. Crackles sounds (CSs) are transient sound biomarkers, with most of the energy ranges from 100 Hz to 2000 Hz [15]. CS can be classified as coarse or fine. Coarse crackles (CSCs) last less than 15 ms and show a low pitch while fine crackles (CSFs) last less than 5 ms with a high pitch [14]. Focusing on respiratory pathologies, CSCs are primarily associated to chronic bronchitis, bronchiectasis and COPD while CSFs are related to lung fibrosis and pneumonia [14]. Figure 1 shows several examples of spectrograms, captured by means of auscultation, from normal and abnormal respiratory sounds.

According to the literature focused on biomedical respiratory sounds, the tasks of detection and classification of adventitious respiratory sounds have been addressed using many approaches that combine signal processing and/or AI, such as Tonal index [16,17], Mel-Frequency Cepstral Coefficients (MFCC) [18,19], Hidden Markov Model (HMM) [20–22], chroma features [23], fractal dimension filtering [24–26], Empirical Mode Decomposition (EMD) [27,28], Gaussian Mixture Models (GMM) [29–31], spectrogram analysis [13,15,32–36], Auto-Regressive (AR) models [37–39], entropy [40–42], wavelet [43–49], Support Vector Machines (SVM) [50–53], Independent Component Analysis (ICA) [54] and Non-negative Matrix Factorization (NMF) [55–59]. Until recent times, the classification of respiratory sound signals has presented a significant challenge, primarily due to the scarcity of clinical respiratory data. This scarcity arises from the laborious and resource-intensive nature of the process required to obtain and label respiratory sounds. However, the previous challenge was mitigated in 2019 with the emergence of the largest public database, the International Conference on Biomedical and Health Informatics (ICBHI) [60,61]. Therefore, research focused on different machine learning approaches has recently increased dramatically, such as Recurrent Neural Networks (RNN) [62], hybrid neural networks [63–67] and above all Convolutional Neural Networks (CNN) [64,68–99]. Thus, the use of these types of deep learning architectures provided promising performance improvements due to their ability that they are able to learn behaviour, both in time and frequency, from large datasets, eliminating the engineer intervention in feature extraction techniques, which reduces the likelihood of human error [100]. Rocha et al. [28] proposed wheezing segmentation by means of harmonic–percussive sound separation, enhancing the harmonic source (associated to the wheezes) and applying EMD. In [48], a particular wavelet transform was proposed to determine the type of harmonic energy distribution contained in wheezes. A novel hybrid neural model was performed in [65,101] that combined CNN and Long Short-Term Memory (LSTM) architecture that operated on the Short-Time Fourier Transform (STFT) spectrogram. In this manner, the CNN architecture extracted the most important features from the spectrogram, while the LSTM model recognized and stored the long-term relationships between these features. Ma et al. [80] demonstrated the promising adventitious sound classification, in the ICBHI database, adding a non-local layer in ResNet architecture with a mix-up data augmentation technique. Demir et al. [78] detailed a pre-trained CNN architecture to classify AS including a set of pooling layers that operated in parallel to fed into a Linear Discriminant Analysis (LDA) classifier that utilized the Random Subspace Ensembles (RSE) method. Nguyen and Pernkopf [92] exploited several transfer learning methods, in which the pre-trained ResNet architectures were used applying several fine-tuning methods with the aim of enhancing the robustness in the respiratory sound classification.

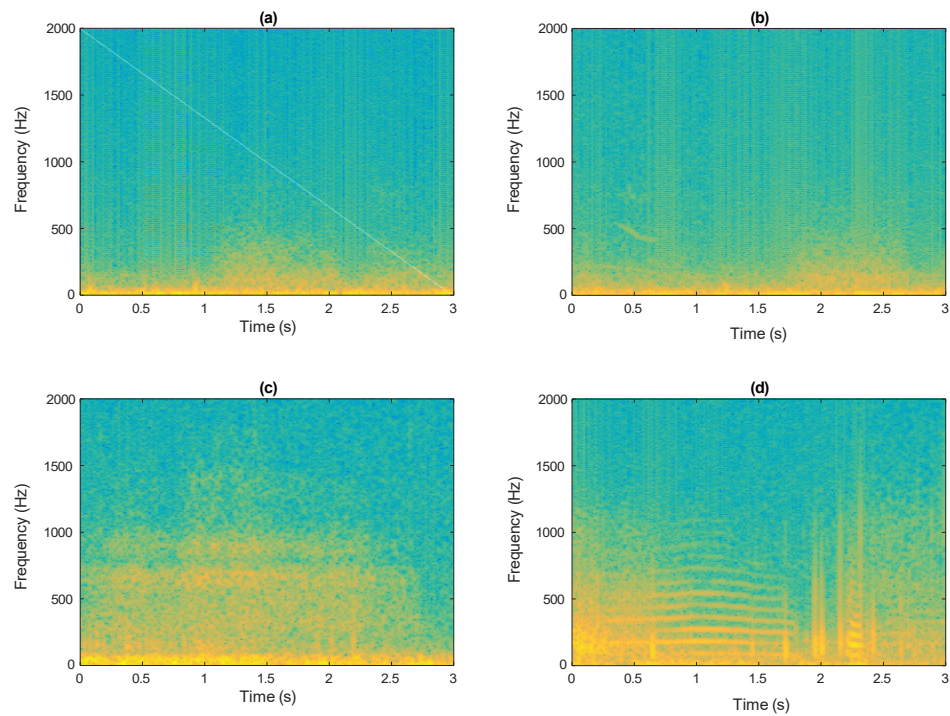

**Figure 1.** Time-frequency representation (spectrogram) of a 3 s auscultation. Normal respiratory sounds (**a**), normal respiratory sounds plus wheezes (**b**), normal respiratory sounds plus crackles (**c**); and finally, normal respiratory sounds plus wheezes plus crackles (**d**).

The study conducted by Rocha et al. [102] highlighted that despite the widespread usage of CNNs as cutting-edge solutions in various research fields, respiratory sound classification can still be improved by exploring alternative time–frequency representations and novel deep learning architectures. In line with the first aspect (time–frequency representations), our recent work [103] performed an extensive analysis of several time–frequency representations and proposed the cochleogram as a suitable representation to capture the unique spectro-temporal features of most adventitious sounds. Our results demonstrated the efficacy of the cochleogram when used as a time-frequency representation in conventional CNN-based architectures in detecting the presence of respiratory abnormalities and classifying these abnormalities focused on wheezing and crackles events. With the second aspect, deep learning architectures, and extending our previously mentioned work, the main contribution of this study is to investigate the performance when the input data, represented by cochleogram, is used to feed the Vision Transformer (ViT) architecture [104] since this input–classifier combination is the first time it has been applied to adventitious sound classification to our knowledge. The main concept behind ViT is to treat the audio signal as a series of spectrogram patches. These patches are then processed by the architecture, which uses a self-attention mechanism to analyze them. By doing so, the ViT model can identify important patterns and connections in the audio signal, enabling it to extract relevant features for classification purposes. Although this approach is still in its early stages, it holds great promise for improving the accuracy and efficiency of audio classification tasks [105,106]. Specifically, this work extends the previous Vision Transformer (ViT) architecture but applied in the context of respiratory sound classification feeding the learning architecture by means of the cochleogram instead of conventional time-frequency representations as occurs in [107].

The structure of this study is performed as follows. The set of time-frequency representations are detailed in Section 2 and the proposed Transformer-based architecture is described in Section 3. The dataset, the metrics, and the compared algorithms are explained in Section 4. Section 5 reports the classification performance of the baseline and other state-of-the-art CNN approaches. Section 6 draws the most relevant conclusions and outlines future work.

**2. Proposed Methodology**

This work proposes a typical classification scheme composed of a feature extraction process followed by a deep learning-based classifier. In particular, several representations are studied including the linear frequency scale STFT and the log-frequency scale MFCC, CQT, and the cochleogram. Then, the classifier based on the Transformer architecture is presented together with the details about the training procedure.

*2.1. Feature Extraction*

In this section, we review the most used TF representations in the literature and introduce the cochleogram, which has recently been successfully applied for classifying adventitious sounds with promising results [103] using state-of-the-art deep learning-based approaches.

2.1.1. Short-Time Frequency Transform (STFT)

Short-Time Frequency Transform (STFT) is a widely utilized time-frequency (TF) representation. For a each $k$-th frequency bin and $m$-th time frame, the STFT coefficients $X(k, m)$ are estimated as

$$X(k,m) = \sum_{n=0}^{N-1} x((m-1) \cdot J + n)w(n)e^{-j\frac{2\pi}{N}kn}, \qquad (1)$$

where $x(n)$ is the input signal, $w(n)$ is the analysis window of length $N$, and $J$ is the time shift expressed in samples. Typically, only the magnitude spectrogram $\mathbf{X} = |\mathbf{X}| \in \mathbb{R}_+^{K \times L}$ is used for analyzing the spectral content, and the phase information is ignored. Nevertheless, STFT spectrograms might not be optimal for examining respiratory sounds as they offer a consistent bandwidth, leading to diminished resolution at lower frequencies, where a significant portion of relevant respiratory spectral content is situated. Additionally, STFT may perform poorly when analysing respiratory sounds in noisy environments [108].

### 2.1.2. Mel-Frequency Cepstral Coefficients (MFCC)

MFCC (Mel-Frequency Cepstral Coefficients) is a commonly used feature extraction technique in speech and audio signal processing. The Mel-frequency scaling is applied to the power spectrum. This involves mapping the frequency axis from the linear scale to a non-linear Mel scale. The Mel scale is a perceptually based scale that more closely represents how humans perceive sound. The Mel scale is defined as follows,

$$\text{Mel}(f) = 1127 \cdot \ln(1 + f/700) \tag{2}$$

where $f$ is the frequency in Hz. The Mel-scaled power spectrum is then passed through a filterbank $H_m(k)$ composed of $M$ triangular filters that mimic the frequency selectivity of the human auditory system as,

$$H_m(k) = \sum_{i=1}^{M} |X(k)|^2 \cdot H_m^i(k), \tag{3}$$

where $|X(k)|$ is the magnitude of the Fourier transform at frequency $k$, and $H_m^i(k)$ is the $i$-th triangular filter in the Mel filter bank centred at Mel frequency $m_i$. The filters are spaced uniformly on the Mel scale (see Equation (2)), and their bandwidths increase with increasing frequency. The output of each filter is then squared and summed over frequency to obtain a measure of the energy in each filter as,

$$S_m = \log\left(\sum_{k=1}^{K} |X(k)|^2 \cdot H_m(k)\right), \tag{4}$$

where $K$ is the number of frequency bins in the Fourier transform. Finally, the discrete cosine transform (DCT) is utilized to decorrelate the filterbank energies and produces a set of coefficients that are often used as features for machine learning architectures as,

$$Y_n = \sqrt{\frac{2}{M}} \sum_{m=0}^{M-1} S_m \cos\left(\frac{\pi n}{M}\left(m + \frac{1}{2}\right)\right), \tag{5}$$

where $S_m$ is the logarithmic scaling of the $m$-th filter bank output and $Y_n$ is the $n$-th MFCC coefficient. The first few MFCC coefficients tend to capture the spectral envelope or shape of the signal, while the higher coefficients capture finer spectral details. The number of MFCC coefficients is typically chosen based on the application, and can range from a few to several dozen.

### 2.1.3. Constant-Q Transform (CQT)

The Constant-Q Transform (CQT) is a type of frequency-domain analysis that is particularly useful for analysing signals that have a non-uniform frequency content, such as musical signals. The CQT is similar to the Short-Time Fourier Transform (STFT), but instead of using a linear frequency scale, it uses a logarithmic frequency scale that is more similar to the frequency resolution of the human auditory system. This logarithmic frequency resolution allows for a more accurate representation of respiratory sounds, which often have a non-uniform frequency content. For an input signal $x(n)$, the CQT can be defined mathematically as:

$$X(k,n) = \sum_{j=n-\lfloor N_K/2 \rfloor}^{j=n+\lfloor N_K/2 \rfloor} x(j) a^*(j - n - N_k/2), \tag{6}$$

where $k$ is the frequency index in the CQT domain, $\lfloor \cdot \rfloor$ denotes towards negative infinity and $a^*(n)$ are the time–frequency atoms defined by

$$a_k(n) = \frac{1}{N_k} w\left(\frac{n}{N_K}\right) \exp\left[-i2\pi n \frac{f_k}{f_s}\right], \tag{7}$$

where $f_k$ is the centre frequency of bin $k$, $f_s$ is the sampling rate, $w(n)$ is the window function (e.g., Hann or Blackman Harris). The window lengths $N_k \in \mathbb{R}$ are inversely proportional to $f_k$ in order to have the same Q-factor for all the frequency bins $k$. The Q factor of bin $k$ is given by

$$Q_k = \frac{f_k}{\Delta f_k}, \tag{8}$$

where $\Delta f_k$ denotes the $-3$ dB bandwidth of the frequency response of the atom $a_k(n)$ and the range of $f_k$ obey

$$f_k = f_1 2^{\frac{k-1}{b}} \tag{9}$$

where $f_1$ is the centre frequency of the lowest frequency bin and $b$ is the number of bins per octave. In fact, the parameter $b$ determines the time-frequency resolution trade-off of the CQT.

Unlike the STFT, where the frequency resolution is constant across all frequency bins, the CQT has a higher frequency resolution for lower frequencies and a lower frequency resolution for higher frequencies. Another advantage of the CQT is that it can provide better time-frequency resolution compared to the STFT for signals with rapidly changing frequencies.

2.1.4. Cochleogram

The gammatone filter is specifically designed to emulate the behaviour of the human cochlea by incorporating non-uniform spectral resolution (i.e., this Cochleogram assigns broader frequency bandwidths to higher frequencies). This adaptable resolution results in a time-frequency (TF) representation robust against noise and acoustic variations [108–110]. In the computation of the cochleogram, a gammatone filter bank is employed. The gammatone filter's impulse response, represented as $g(t)$, is obtained by multiplying a gamma distribution and a sinusoidal function as follows:

$$g(t) = t^{o-1} e^{2\pi b(f_c) t} \cos(2\pi f_c t), \ t > 0 \tag{10}$$

where the filter's bandwidth is determined by both the filter order $o$ and the exponential decay coefficient $b(f_c)$ associated with the centre frequency $f_c$ of the filter in Hertz. The centre frequencies are evenly distributed along the equivalent rectangular bandwidth (ERB) scale as,

$$b(f_c) = 1.019 \cdot \text{ERB}(f_c) \tag{11}$$

$$\text{ERB}(f_c) = 24.7 \cdot \left(4.37 \cdot \frac{f_c}{1000} + 1\right) \tag{12}$$

Following the application of the gammatone filter to the signal, as detailed in [110], a representation akin to the spectrogram is generated by summing the energy in the windowed signal for each frequency channel. This process can be expressed as follows:

$$C(k, m) = \sum_{n=0}^{N-1} |\hat{X}(k, n)| w(n), \tag{13}$$

where $\hat{X}(k, n)$ is the gammatone filtered signal, $k = 1, \ldots, K$ is the number of gammatone filters and $C(k, m)$ represents the coefficient corresponding to the centre frequency $f_c(k)$ for the $m$-th frame and $w(n)$ refers to the windowed signal. In this work, we used $K = 64$ gammatone filters with the central frequencies $f_c(k)$ uniformly distributed between 100 Hz and $\frac{f_s}{2}$ Hz, respectively. Note that most adventitious respiratory sounds, particularly wheezing and crackles, exhibit predominant content in this spectral range. In this paper, we use an order $o = 4$ as it yields satisfactory results in emulating the human auditory filter [108].

As an example, Figure 2 shows a comparison of TF representations computed by means of STFT, Mel-scaled spectrogram, CQT, and cochleogram. Among these representations, the cochleogram stands out for its ability to provide a highly accurate depiction of adventitious sounds. In fact, this gammatone filtering technique with non-uniform resolution proves to be particularly effective in modelling the low spectral respiratory content.

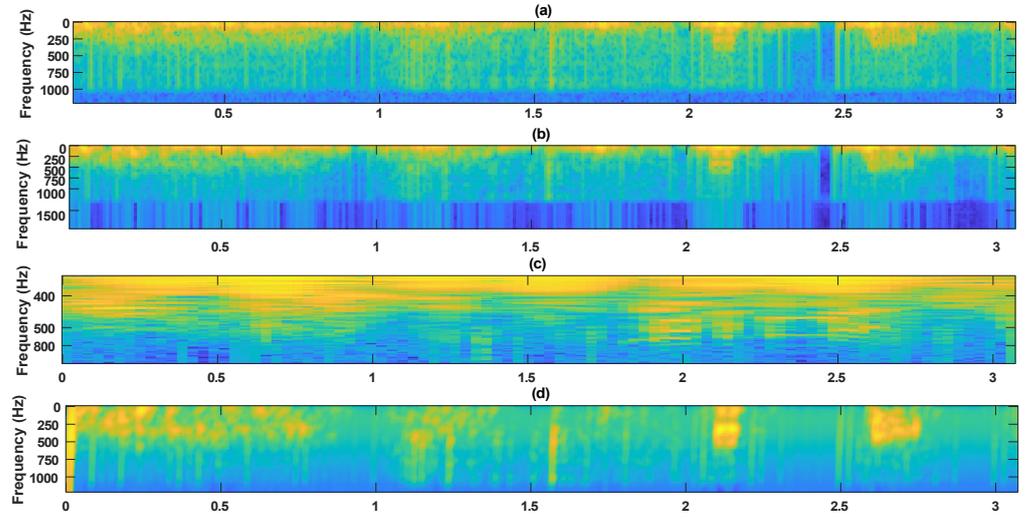

**Figure 2.** Magnitude spectrogram representations of a 3.2 s respiratory cycle from ICBHI [60] presenting a wheezing event in the interval [2.6–2.65] s. STFT spectrogram (**a**), Mel-scaled spectrogram (**b**), Constant-Q (**c**) and cochleogram (**d**).

## 3. Vision Transformer-Based Classifier

Competitive neural sequence transduction models typically follow an encoder–decoder framework. Within this structure, the encoder processes an input sequence of symbol representations $(x_1, \ldots, x_n)$ and transforms it into a sequence of continuous representations $z = (z_1, \ldots, z_n)$. Subsequently, the decoder utilizes the obtained continuous representations $z$ to iteratively generate an output sequence $(y_1, \ldots, y_m)$ of symbols. The model operates in an auto-regressive manner at each step, incorporating previously generated symbols as additional input for generating the next symbol.

The Transformer model adopts this overarching architecture, employing stacked self-attention and point-wise, fully connected layers for both the encoder and decoder. Figure 3 illustrates these components in the left and right halves, respectively.

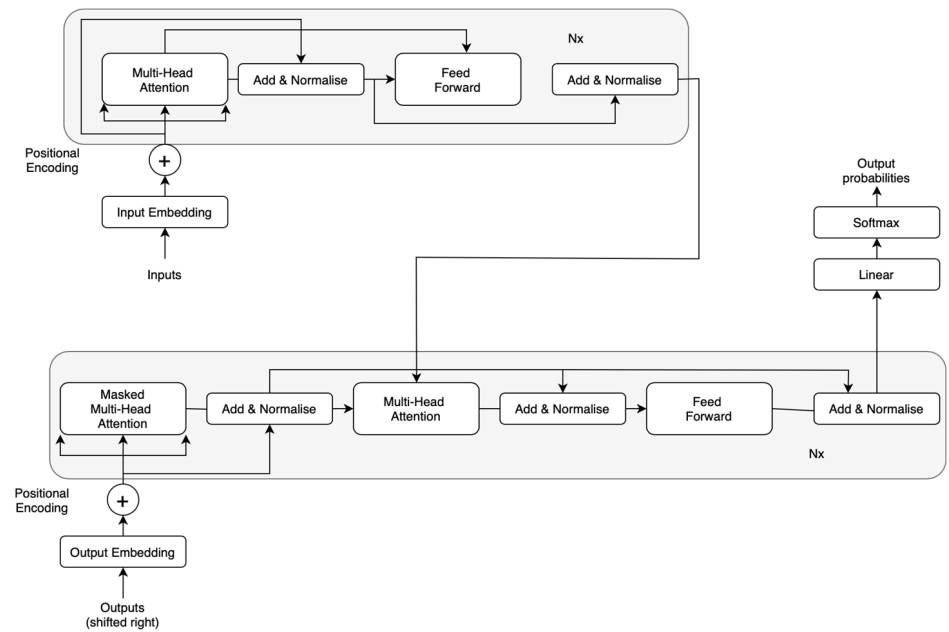

**Figure 3.** Architecture based on the Vision Transformer model [104].

Our model closely follows the original Vision Transformer architecture (ViT) [104]. It begins by partitioning an image into predetermined-sized patches, followed by linear embedding of each patch. Position embedding is then applied, and the resulting sequence of vectors is fed into a conventional transformer encoder. To facilitate classification, we adopt the standard technique of including a supplementary learnable "classification token" in the sequence, as illustrated in Figure 3. While this approach shares similarities with the original transformer [111] used in natural language processing tasks, it is specifically tailored to handle images. To manage computational costs, ViT computes relationships among pixels within small, fixed-sized patches of the image. These patches undergo linear embedding, and position embeddings are added. The resulting vector sequence passes through a standard transformer encoder consisting of a stack of $N$ = six identical layers, each comprising two sub-layers. The first sub-layer employs a multi-head self-attention mechanism, while the second sub-layer is a simple, position-wise fully connected feed-forward network. To facilitate information flow and aid gradient propagation, we introduce a residual connection around each sub-layer, followed by layer normalization. This implies that the output of each sub-layer is computed through LayerNorm(x + Sublayer(x)), where Sublayer(x) represents the function implemented by the sub-layer itself. To maintain these residual connections, all sub-layers in the model, including the embedding layers, produce outputs of dimension dmodel = 512. The decoder is constructed with a stack of $N$ = 6 identical layers. Each decoder layer comprises two sub-layers, similar to the encoder. However, the decoder introduces an additional third sub-layer, performing multi-head attention over the output of the encoder stack. Similar to the encoder, residual connections surround each sub-layer, followed by layer normalization.

## 4. Materials and Methods

This section describes the dataset, the classification metrics and several deep learning architectures successfully applied in this biomedical context.

*4.1. Dataset*

In this work, the publicly available ICBHI 2017 database [61] is used. It consists of 920 recordings captured using variable durations and sampling rates from 126 patients. The set of recordings were acquired from different chest points by means of various equipment, such as AKG C417L Microphone (AKGC417L), 3M Littmann Classic II SE Stethoscope (LittC2SE), 3M Litmmann 3200 Electronic Stethoscope (Litt3200), and WelchAllyn Meditron Master Elite Electronic Stethoscope (Meditron). However, each recording has been down-sampled to 4 kHz, since the respiratory sounds to analyze do not exceed 2 kHz [10,56,112]. Each recording is temporally labelled, specifying the start and end of each respiratory cycle, and indicating the presence or absence of adventitious sounds, such as crackles, wheezes, or both. Specifically, we have adjusted to 6 s each respiratory cycle due to the findings that show better performance for this duration of time [88]. Table 1 provides a summary of the number of cycles per class: normal (absence of crackles and wheezes), crackles, wheezes, or both of them. More details about the ICBHI database can be found at [61].

**Table 1.** Overview, focusing on the type and number of respiratory cycles, of the ICBHI 2017 dataset.

| Type of Respiratory Cycle | Number of Respiratory Cycles |
|---|---|
| Crackle | 1.864 |
| Wheeze | 886 |
| Crackle + Wheeze | 506 |
| Normal | 3.642 |
| **Total** | **6.898** |

*4.2. Metrics*

A suite of metrics was employed to evaluate the performance of the proposed method through an analysis of the classification confusion matrix. These metrics encompassed Accuracy (*Acc*), Sensitivity (*Sen*), Specificity (*Spe*), Precision (*Prec*), and Score (*Sco*). For each fold, a confusion matrix was generated, and the metrics were computed using the aggregated confusion matrix from the 10-fold cross-validation. The metrics were defined based on the following parameters: true positive (*TP*) representing adventitious sounds correctly classified, true negative (*TN*) indicating normal respiratory sounds (healthy sound) correctly classified, false positives (*FP*) denoting normal respiratory sounds incorrectly classified as adventitious sounds, and false negatives (*FN*) representing adventitious sounds incorrectly classified as normal respiratory sounds.

- Accuracy (*Acc*) measures the number of correctly classified adventitious sounds and normal respiratory sounds cycles from the total number of test samples.

$$Acc = \frac{TP + TN}{TP + TN + FP + FN} \quad (14)$$

- Sensitivity (*Sen*) is defined as the number of correctly detected adventitious sounds class from the total number of predicted adventitious sound events.

$$Sen = \frac{TP}{TP + FN} \quad (15)$$

- Precision (*Pre*) is defined as the positive predictive value (PPV) where a true positive is considered as the target event, when the test makes a positive forecast, and the subject has a positive result.

$$Prec = \frac{TP}{TP + FP} \quad (16)$$

- Specificity (*Spe*) represents the correctly labelled normal respiratory sound events (*TN*) from the total number of normal respiratory sound events (*TN* + *FP*).

$$Spe = \frac{TN}{TN + FP} \quad (17)$$

- Score (*Sco*) represents a general measure of the quality of the classifier as an average of the sensitivity and specificity metrics.

$$Sco = \frac{Sen + Spe}{2} \quad (18)$$

*4.3. Compared State-of-the-Art Architectures*

Some of the most used state-of-the-art learning models, applied in the classification of adventitious sound events, are described below.

**Baseline CNN** [102]. The approach outlined here, proposed by the creators of the ICBHI dataset, serves as the baseline method for assessment. The architectural design of the method involves two convolutional layers followed by a deep neural network (DNN) layer utilizing leaky ReLU activation functions and a softmax output function. For training the deep learning models, the Adam optimization algorithm was applied with a learning rate of 0.001 and a batch size of 16 over a span of 30 epochs. To counteract overfitting, an early stopping technique was employed, wherein the training procedure halted if the validation loss did not improve by more than 25% of the training set for a consecutive streak of 10 epochs.

**AlexNet** [113] was the initial Convolutional Neural Network (CNN) to secure victory in the ImageNet Large Scale Visual Recognition Challenge (ILSVRC) was documented by Russakovsky et al. [114]. This groundbreaking model, known as AlexNet, featured five convolutional layers, three pooling layers, and three fully connected layers, showcasing the efficacy of deep CNNs and the ReLU activation function. Despite its successes, AlexNet faced significant drawbacks, including high computational costs during training and a substantial number of parameters that heightened the risk of overfitting. For the classification of adventitious sounds, the architecture outlined in [115] was adopted. The obtained results demonstrated an accuracy (*Acc*) of 83.78% for normal breathing sound detection, 83.78% for crackle sound detection, and 91.89% for wheeze sound detection.

**VGG16** [116] constitutes a deep architecture comprising 16 convolutional layers, five pooling layers, and three fully connected layers. The significance of small filter sizes and the utilization of max-pooling layers for down-sampling is emphasized in VGG16. However, the architecture entails a high computational cost due to its numerous parameters, rendering it challenging to train on resource-constrained devices or within limited memory constraints for real-time applications. Moreover, VGG16 is susceptible to overfitting given its abundance of parameters, potentially resulting in suboptimal generalization performance and restricting its adaptability to new and unseen data. This architectural framework has been employed for adventitious sound classification in [117], achieving an accuracy (*Acc*) of 62.5% for a 4-class classification (i.e., healthy, wheezing, crackle, and both wheezing+crackle).

**ResNet50** [118] introduced the innovative concept of residual connections, allowing the network to learn residual functions rather than directly mapping the underlying features. The significance of depth and residual connections in CNNs was exemplified by ResNet50, a variant with 50 layers. However, ResNet50 is not without limitations, particularly in terms of memory requirements and interpretability of learned features. The extensive layer count in ResNet50 translates to substantial memory demands, necessitating a considerable amount of storage for activations and gradients during the training phase. Additionally, the surplus of layers hampers the interpretation of acquired features, a critical aspect for feature selection and transfer learning tasks. These challenges constrain the ability to gain insights into the intrinsic patterns and structure of the data. This architectural framework has been applied to adventitious sound classification in [117], yielding an accuracy (*Acc*) of 62.29% for a 4-class classification.

In our experiments, we utilized the implementations of AlexNet, VGG16, and ResNet50 as provided by their respective authors. Given that these models were initially designed to handle images as inputs, we converted each computed TF representation matrix into an image format using the Viridis Colour Map. The Viridis Colour Map was chosen for its application of a consistent colour gradient that transitions from blue to green to yellow.

*4.4. Training Procedure*

In our investigation, a diverse set of metrics was incorporated, and the 10-fold cross-validation method was employed to categorize patients into training, testing, and validation sets. This involved dividing the entire patient dataset into 10 equivalent segments or 'folds,' with each fold serving as the testing set in rotation while the remaining 9 folds were utilized for training. This process was iterated 10 times, ensuring that each dataset segment was employed precisely once as the testing set.

The adoption of 10-fold cross-validation guarantees a balanced and thorough training and validation process for the model. Breaking down the data into 10 parts minimizes the potential for biases or anomalies that might arise from simpler splits, such as a 70–30 or 80–20 division.

The same training methodology has been employed for the compared architectures in Table 2. The training process involved a total of 30 epochs, utilizing a batch size of 16, a learning rate set at 0.001, and the adaptive data momentum (ADAM) optimization algorithm.

**Table 2.** A comprehensive overview of several conventional CNN architectures used in this work.

| Architectures | (Conv. Layers) | (Pool Layers) | (Activation) | (Parameters) |
|---|---|---|---|---|
| BaselineCNN | 2 ($5 \times 5, 3 \times 3$) | 2 ($2 \times 2$) | *LeakyReLU* | 8 M |
| AlexNet | 5 ($11 \times 11, 5 \times 5, 3 \times 3$) | 3 ($3 \times 3, 2 \times 2$) | *ReLU* | 160 M |
| VGG16 | 13 ($3 \times 3$) | 5 ($2 \times 2$) | *ReLU* | 138 M |
| ResNet50 | 50 ($7 \times 7, 3 \times 3, 1 \times 1$) | 1 ($3 \times 3$) | *ReLU* | 25.5 M |

The final value for the employed metrics (i.e., accuracy, sensitivity, specificity, etc.) is computed by averaging the individual values for each 10-fold iteration.

The research experiments were conducted utilizing Tensorflow and Keras, which were installed on a computer equipped with an Intel(R) Core(TM) 12th Gen i9-12900, a NVIDIA GeForce RTX3090 GPU, and 128 GB of RAM.

In order to assess the computational implications of the research, Table 3 is provided to elucidate the time breakdown, measured in minutes per epoch, during the training of individual models. The computational cost across various neural network architectures can differ significantly owing to variations in model architecture, depth, and design choices. While deeper architectures generally enhance representation learning, they concurrently escalate computational requirements. In this particular task, the heightened depth of AlexNet results in the highest computational cost. Conversely, VGG exhibits increased computational demands due to its uniform architecture and the incorporation of smaller convolutional filters ($3 \times 3$) across multiple layers. ResNet, characterized by its depth, also incurs a higher computational cost, albeit mitigated by the use of skip connections (residual blocks) addressing the vanishing gradient problem. The BaselineCNN, being relatively shallow, bears a lower computational cost compared to deeper counterparts like AlexNet, ResNet, and VGG, rendering it suitable for simpler tasks or scenarios with restricted computational resources. The computational cost associated with Vision Transformers is variable. While they may necessitate fewer parameters than traditional CNNs for specific tasks, the introduction of the self-attention mechanism introduces additional computational complexity. The selection of an architecture is often contingent on the task's complexity. For more intricate tasks, deeper architectures like Vision Transformers may prove advantageous, exploring novel paradigms that potentially offer competitive performance with reduced computational overhead.

**Table 3.** Comparison of the computational time per epoch of the different architectures.

| Comparison | Time (min)/Epoc |
|---|---|
| AlexNet | 53.2868 |
| ResNet50 | 20.6706 |
| VGG16 | 39.5655 |
| BaselineCNN | 0.4488 |
| ViT | 4.9948 |

## 5. Evaluation

This section assesses the performance of the Vision Transformer (ViT) in the context of adventitious sound classification in comparison to other state-of-the-art methods discussed earlier in Section 4.3. The evaluation of the proposed method occurs in two distinct scenarios, as outlined in Section 4.1, utilizing the ICBHI dataset. In the first scenario, a two-class (binary) classification is conducted to determine the presence of wheezes and crackles in each respiratory cycle. In the second scenario, a four-class classification is undertaken to identify four specific classes: healthy, wheezing, crackles, and both (wheezing + crackles).

### 5.1. 2-Class (Binary) Classification Results

We have evaluated the performance of the ViT with respect to the other state-of-the-art architectures for the task of detecting the presence of crackles and wheezes in respiratory sound signals that correspond to individual respiratory cycles.

Figure 4 presents the accuracy results of the evaluated models using the studied time-frequency representations (STFT, MFCC, CQT, and the cochleogram) as inputs for distinguishing wheezes from other sounds and crackles from other sounds. The compared neural network architectures include BaselineCNN, AlexNet, VGG16 and ResNet50, and the ViT (Vision Transformer). Results indicate that the proposed method, based on the ViT model using the cochleogram, achieved the best performance for both crackle and wheezing classification, with an average accuracy $Acc$ = 85.9% for wheezes and $Acc$ = 75.5% for crackles detection. In fact, employing transformers to capture bi-directional dependencies in COPD audio signals holds potential in predicting adventitious sounds, even in the presence of sparse sound events. It is worth noting that the STFT spectrogram also provided competitive performance, with an accuracy $Acc$ = 82.1% for wheezes and $Acc$ = 72.2% for crackles. These results may be due to the fact that the effective low-pass filtering of the frequencies of interest and the use of an appropriate window length and hop size [103], resulting in the accurate detection of adventitious sound events even when a linear frequency scale is used. Interestingly, the log-scale frequency transforms (MFCC and CQT) clearly underperform, showing $Acc$ = 79.9% for wheezes and $Acc$ = 70.1% for crackles detection in the case of MFCC and $Acc$ = 78.8% for wheezes and $Acc$ = 68.8% using the CQT spectrogram. Although MFCC and CQT are effective for modelling speech and music signals, both do not seem to be the most appropriate TF representation for capturing the most predominant content in the context of adventitious sounds. Comparable behaviour is observed among the state-of-the-art neural network architectures when applied to the task of crackle detection, except for the AlexNet model. In the case of the AlexNet model detecting crackles, the MFCC yield the highest accuracy result, specifically $Acc$ = 67.9%, among the compared TF input representations. However, this accuracy is diminished by approximately 8% when compared AlexNet to the peak performance, $Acc$ = 75.5%, achieved by the proposed method that employs the Vision Transformer (ViT) architecture fed from the cochleogram input. It is also interesting to highly the narrower dispersion of the results obtained by the ViT architecture, which demonstrate the robustness of this method with respect to the different acoustic conditions of the input respiratory cycle. Finally, BaselineCNN and VGG16 outperform the AlexNet and ResNet50 architectures.

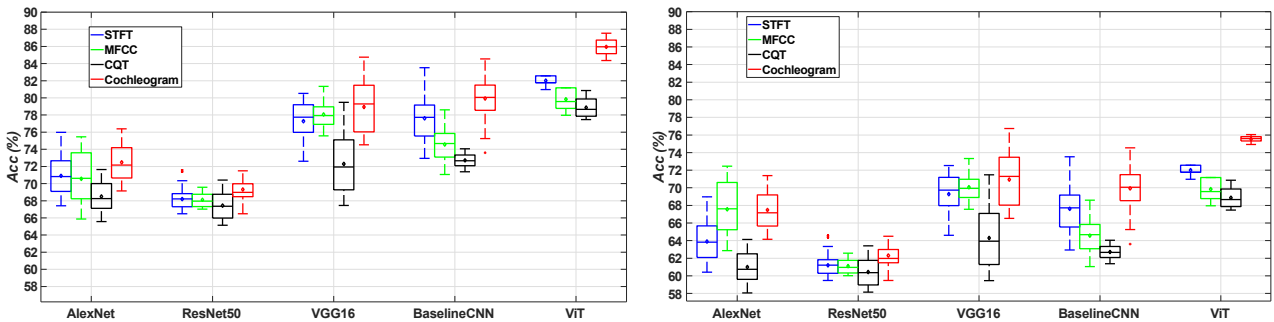

**Figure 4.** The accuracy outcomes for the assessed deep learning architectures, utilizing TF representations for feature extraction, are presented in the task of a 2-class scenario: wheezes (yes/no) on the left side and crackles (yes/no) on the right side within the ICBHI database. Each box in the visualization corresponds to 50 data points, with each point associated with a 10-fold cross-validation. The median value for each box is represented by a line in the middle, while the lower and upper lines indicate the first and third quartiles. The average value is denoted by a diamond shape at the centre of each box. The range of the remaining samples, excluding outliers, is depicted by the lines extending above and below each box. Outliers, defined as data points exceeding 1.5 times the interquartile range from the sample median, are marked with crosses.

To determine the statistical significance of the findings presented in Figure 4, two widely referenced and robust non-parametric tests have been used, specifically, the Mann–Whitney U test and the Wilcoxon signed-rank test [119,120]. In particular, Tables 4 and 5 display the results of these tests, which compared the classification performance of the ViT, VGG16, BaselineCNN, AlexNet, and ResNet50 models using cochleogram representation for both crackles and wheezes. The null hypothesis $H_o$ for these tests assumes that there is no significant difference between the two distributions being compared, while the alternative hypothesis $H_1$ postulates that a significant difference exists. The $p$-value, which indicates the probability of obtaining results as extreme as those observed assuming that $H_o$ is true, determines whether we reject or accept $H_o$. Our analysis, using a significance level of $\alpha = 0.05$, reveals that the $p$-value for all cases did not exceed the significance level, allowing us to reject $H_o$ and conclude that the ViT performs significantly better than VGG16, BaselineCNN, AlexNet, and ResNet50 for both crackles and wheezes classification.

**Table 4.** The Mann–Whitney U test and Wilcoxon signed-rank test were performed on the data sets shown in Figure 4 (left side) with a significance level of $\alpha = 0.05$.

| Comparison | Mann-Whitney U Test | Wilcoxon Signed-Rank Test | Significantly Better |
|---|---|---|---|
| **Cochleogram** | (*p*-Value) | (*p*-Value) | (Yes/No) |
| ViT vs. VGG16 | $4.99 \times 10^{-18}$ | $6.13 \times 10^{-13}$ | yes |
| ViT vs. BaselineCNN | $5.34 \times 10^{-18}$ | $8.61 \times 10^{-15}$ | yes |
| ViT vs. AlexNet | $6.91 \times 10^{-18}$ | $5.68 \times 10^{-15}$ | yes |
| ViT vs. ResNet50 | $8.16 \times 10^{-18}$ | $6.84 \times 10^{-15}$ | yes |

**Table 5.** The Mann–Whitney U test and Wilcoxon signed-rank test were performed on the data sets shown in Figure 4 (right side) with a significance level of $\alpha = 0.05$.

| Comparison | Mann-Whitney U Test | Wilcoxon Signed-Rank Test | Significantly Better |
|---|---|---|---|
| **Cochleogram** | (*p*-Value) | (*p*-Value) | (Yes/No) |
| ViT vs. VGG16 | $9.28 \times 10^{-18}$ | $1.77 \times 10^{-15}$ | yes |
| ViT vs. BaselineCNN | $2.74 \times 10^{-17}$ | $1.64 \times 10^{-14}$ | yes |
| ViT vs. AlexNet | $6.35 \times 10^{-18}$ | $1.43 \times 10^{-13}$ | yes |
| ViT vs. ResNet50 | $5.10 \times 10^{-18}$ | $1.14 \times 10^{-11}$ | yes |

In this work, the accuracy (*Acc*) has been used as the main metric to provide a general measure of the classification performance, taking into account both successful adventitious events (*TP* and *TN*) as well as false adventitious events (*FP*) and undetected adventitious events (*FN*). In order to compare the proposed method with other state-of-the-art algorithms, Table 6 shows other metrics, such as Sensitivity (*Sen*), Specificity (*Spe*), Score (*Sco*), and Precision (*Pre*), which have also been proposed in the literature to assess the performance associated to the adventitious sound classification. The results show that using the ViT architecture with cochleogram as input gives the best classification performance for each type of adventitious sound. Compared to using STFT, using cochleogram improves wheezes' classification by about 4.1% and crackles' classification by about 2.3%, on average. STFT is ranked in second place in terms of performance, followed by MFCC and CQT, which rank last. Specifically, STFT outperforms MFCC by at least 2.1% on average for both wheezes and crackles. Furthermore, MFCC outperforms CQT with a minimum average improvement of 2.0% for both wheezes and crackles. Focusing on the behaviour of the compared systems based on each metric, the highest values are obtained in terms of Specificity (*Spe*), reporting that the architectures are capable of accurately predicting when patients are healthy. However, the underperformance is shown in terms of Sensitivity (*Sen*) and Precision (*Pre*) since the lowest values are related to the Precision (*Pre*) for all evaluated neural network architectures. This fact reveals that the number of false positives

(healthy patients classified as sick) exceeds the number of false negatives (sick patients classified as healthy). Nevertheless, this outcome can be considered highly advantageous from a medical standpoint, since it provides assurance that patients who exhibit even the slightest doubt or uncertainty concerning the presence of a respiratory disease will receive immediate attention and the necessary medical care they require.

**Table 6.** Sensibility *Sen*, specificity *Spe*, score *Sco* and precision *Pre* results for the proposed method and the other evaluated neural network architectures applying different TF representations for the task of binary 2-class scenario crackles (yes/no) in the ICBHI database. The maximum value for each metric is highlighted in bold.

| Model | TF | Sensibility (*Sen*) | | Specificity (*Spe*) | | Score (*Sco*) | | Precision (*Pre*) | |
|---|---|---|---|---|---|---|---|---|---|
| | | Wheezes | Crackles | Wheezes | Crackles | Wheezes | Crackles | Wheezes | Crackles |
| AlexNet | STFT | 65.1 | 55.1 | 70.1 | 60.1 | 67.6 | 57.6 | 44.8 | 44.8 |
| | MFCC | 62.8 | 52.8 | 68.9 | 59.9 | 65.8 | 56.3 | 39.5 | 39.5 |
| | CQT | 61.7 | 51.7 | 68.3 | 55.3 | 65.0 | 53.5 | 37.6 | 37.6 |
| | Cochleogram | 66.3 | 55.1 | 72.1 | 62.7 | 69.2 | 58.9 | 44.4 | 44.4 |
| ResNet50 | STFT | 61.7 | 51.7 | 72.1 | 62.1 | 66.9 | 56.9 | 38.4 | 38.4 |
| | MFCC | 59.0 | 49.0 | 69.1 | 59.0 | 64.0 | 54.0 | 38.1 | 38.1 |
| | CQT | 58.4 | 48.4 | 69.0 | 59.0 | 64.7 | 53.7 | 36.8 | 36.8 |
| | Cochleogram | 62.2 | 51.7 | 71.7 | 61.7 | 66.9 | 56.7 | 39.4 | 39.4 |
| VGG16 | STFT | 69.4 | 59.4 | 81.9 | 71.9 | 75.6 | 65.6 | 46.4 | 46.4 |
| | MFCC | 62.7 | 52.7 | 76.5 | 66.5 | 69.6 | 59.6 | 44.8 | 44.8 |
| | CQT | 59.0 | 49.0 | 72.6 | 62.6 | 65.8 | 66.8 | 36.4 | 36.4 |
| | Cochleogram | 71.6 | 59.4 | 82.7 | 72.7 | 77.1 | 66.0 | 48.9 | 48.9 |
| BaselineCNN | STFT | 66.7 | 61.7 | 82.4 | 72.4 | 74.5 | 67.0 | 46.3 | 46.3 |
| | MFCC | 62.8 | 56.8 | 80.3 | 70.3 | 71.58 | 63.5 | 44.9 | 44.9 |
| | CQT | 60.8 | 53.8 | 78.0 | 68.0 | 69.4 | 60.9 | 38.3 | 38.3 |
| | Cochleogram | 67.7 | 62.8 | 85.8 | 75.8 | 76.7 | 65.3 | 50.3 | 50.3 |
| ViT | STFT | 71.9 | 62.9 | 85.0 | 75.0 | 78.5 | 69.0 | 52.4 | 52.4 |
| | MFCC | 67.9 | 59.9 | 82.9 | 72.9 | 75.4 | 66.4 | 50.3 | 50.3 |
| | CQT | 65.9 | 57.9 | 80.5 | 70.5 | 73.2 | 64.2 | 47.7 | 47.7 |
| | Cochleogram | **76.0** | **65.2** | **91.0** | **80.2** | **83.5** | **71.7** | **57.6** | **57.6** |

### 5.2. 4-Class Classification Results

We assessed the performance of the Vision Transformer (ViT) in comparison to other state-of-the-art architectures within a multiclass classification scenario. The goal was to distinguish between normal respiratory sounds (healthy) and respiratory sounds featuring any type of the following adventitious sounds: crackles, wheezes, or both crackles and wheezes in an input breathing cycle.

Figure 5 displays the obtained results in terms of accuracy (*Acc*). As can be seen, the ViT architecture using the cochleogram input TF representation outperforms all the compared methods (*Acc* = 67.9%). Similar to the 2-class scenario, using the STFT provides better results than the other log-scale transforms (MFCC and CQT). Identical behaviour can be observed for all the compared TF representations, independently of the evaluated architecture. VGG16 and BaselineCNN using the cochleogram obtained competitive results (*Acc* = 63.9% for wheezes and 62.8% for crackles), and clearly outperform the results using the AlexNet and ResNet50 architectures.

To establish the statistical significance of the performance of the ViT architecture in comparison to the other assessed architectures, we followed the same procedure outlined in the 2-classes scenario (refer to Section 5.1, in Tables 4 and 5). Specifically, the results of these tests are presented in Table 7, signifying that the ViT architecture brings a significant enhancement to the 4-class classification of respiratory sounds when compared to the other neural network architectures under evaluation.

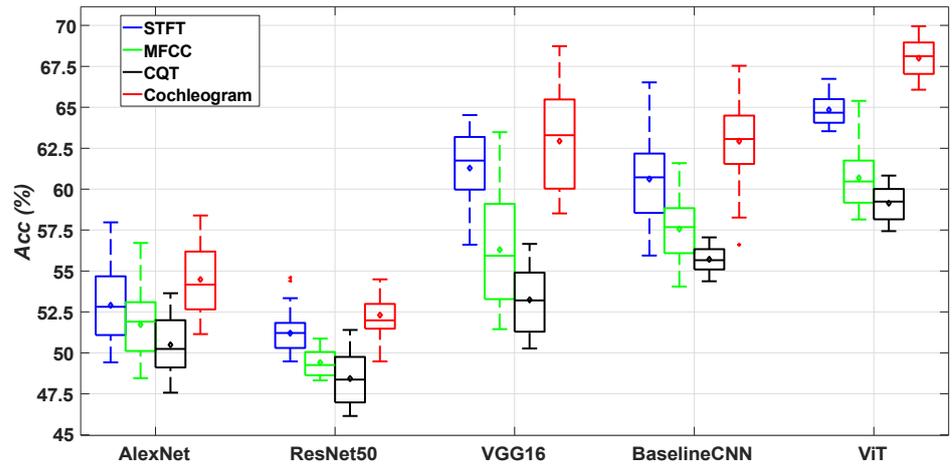

**Figure 5.** The accuracy outcomes for the assessed deep learning architectures, employing TF representations for feature extraction, are presented in the context of a 4-class scenario (normal, wheezes, crackles, wheezes + crackles) within the ICBHI database. Each box in the visualization corresponds to 50 data points, with each point associated with a 10-fold cross-validation. The median value for each box is depicted by a line in the middle, while the lower and upper lines indicate the first and third quartiles. The average value is represented by a diamond shape at the centre of each box. The range of the remaining samples, excluding outliers, is portrayed by the lines extending above and below each box. Outliers, defined as data points exceeding 1.5 times the interquartile range from the sample median, are marked with crosses.

**Table 7.** The Mann–Whitney U test and Wilcoxon signed-rank test were performed on the data sets shown in Figure 5 with a significance level of $\alpha$ = 0.05.

| Comparison | Mann-Whitney U Test | Wilcoxon Signed-Rank Test | Significantly Better |
| --- | --- | --- | --- |
| **Cochleogram** | (*p*-Value) | (*p*-Value) | (Yes/No) |
| ViT vs. VGG16 | $5.46 \times 10^{-15}$ | $3.67 \times 10^{-13}$ | yes |
| ViT vs. BaselineCNN | $5.61 \times 10^{-16}$ | $3.55 \times 10^{-15}$ | yes |
| ViT vs. AlexNet | $6.91 \times 10^{-18}$ | $1.77 \times 10^{-15}$ | yes |
| ViT vs. ResNet50 | $6.59 \times 10^{-18}$ | $2.77 \times 10^{-15}$ | yes |

Although in this paper we have selected accuracy (*Acc*) as the main metric, the metrics of sensitivity, specificity, score, and precision have been included to provide a more comprehensive analysis of the performance of the proposed method enabling comparison with other state-of-the-art methods as shown in Table 8. Similarly to the two-class scenario, the best results are obtained using the ViT + cochleogram, independently of the compared architecture or input TF representation. In general, it can be observed that the classification performance is better in terms of *Spe* so, the architectures seem to characterize better healthy sounds than adventitious sounds. Moreover, as in the two-class scenario, the false positive (healthy patients classified as sick) remains higher than the false negative (sick patients classified as healthy) and, consequently, *Sen* values are higher than *Pre* values. Moreover, it can be observed that the results in the four-class scenario are worse than in the two-class binary scenario. In fact, all previous metrics may provide lower values when considering the joint occurrence of crackles and wheezes as an independent class. That is, the individual detection of wheezes or crackles when both are present is reported as a prediction error. However, to allow a fair comparison with other methods in the literature, we have used the same metric definitions than in the ICBHI challenge.

**Table 8.** Sensibility, specificity, score, and precision results for the proposed method and the other evaluated neural network architectures applying different TF representations for the task of 4-class scenario in the ICBHI database. The maximum value for each metric is highlighted in bold.

| Model | TF | Sensibility (Sen) | Specificity (Spe) | Score (Sco) | Precision (Pre) |
|---|---|---|---|---|---|
| AlexNet | STFT | 45.7 | 57.1 | 51.4 | 39.8 |
| | MFCC | 42.8 | 57.0 | 49.9 | 35.1 |
| | CQT | 42.8 | 57.0 | 49.4 | 34.3 |
| | Cochleogram | 45.12 | 59.74 | 52.43 | 38.48 |
| ResNet | STFT | 40.4 | 58.1 | 49.2 | 38.4 |
| | MFCC | 39.0 | 55.0 | 47.0 | 32.4 |
| | CQT | 38.4 | 54.0 | 46.2 | 31.8 |
| | Cochleogram | 41.7 | 57.7 | 49.7 | 34.4 |
| VGG16 | STFT | 49.7 | 67.9 | 58.84 | 46.4 |
| | MFCC | 46.7 | 62.5 | 54.6 | 44.8 |
| | CQT | 43.0 | 58.6 | 50.8 | 36.4 |
| | Cochleogram | 53.4 | 68.7 | 61.0 | 48.9 |
| BaselineCNN | STFT | 51.6 | 65.4 | 58.5 | 46.3 |
| | MFCC | 47.8 | 63.3 | 55.5 | 44.9 |
| | CQT | 45.8 | 61.0 | 53.4 | 38.3 |
| | Cochleogram | 52.7 | 68.8 | 60.7 | 50.3 |
| ViT | STFT | 52.9 | 68.0 | 60.5 | 45.4 |
| | MFCC | 49.9 | 64.9 | 57.4 | 42.3 |
| | CQT | 47.9 | 63.5 | 55.7 | 40.7 |
| | Cochleogram | **56.6** | **71.3** | **64.0** | **50.2** |

Table 9 presents a comparison of recent state-of-the-art methods in the literature for classifying adventitious respiratory sounds, focusing on the four-class scenario similar to the ICBHI challenge. It is worth noting that the majority of these methods incorporate Short-Time Fourier Transform (STFT) in the preprocessing step to compute the Time-Frequency (TF) representation of the input data, followed by CNN-based approaches for classification. The reported performance values vary significantly due to differences in the evaluation process, such as the use of specific subsets of the ICBHI database or selected evaluation metrics, making direct comparisons challenging [22,77–79,82,83]. Despite these challenges, the highest reported performance [83] achieves an accuracy (*Acc*) of 80.4%, albeit using a subset of the ICBHI dataset. In terms of the standard ICBHI database metrics (*Sen*, *Spe*, and *Sco*), the Mel+RNN approach [72] achieves the best performance with values of *Sen* = 64.0%, *Spe* = 84.0%, and *Sco* = 74.0%. These results suggest that RNN-based approaches are competitive or even superior to the widely studied CNN-based approaches. It is noteworthy that the ViT can be considered a type of RNN network (Bi-LSTM). Nevertheless, Table 9 underscores that there is still room for improvement in the field of biomedical signal processing and machine learning.

**Table 9.** Comparison of the four-class classification performance (normal vs. wheezes vs. crackles vs. crackles + wheezes) is presented between the proposed method and state-of-the-art approaches using the ICBHI database. The temporal length of the respiratory cycle (RC) was taken into account, incorporating zero padding to ensure a fixed duration for respiratory cycles. The abbreviations used include bi-ResNet (bilinear ResNet), NL (Non-Local), SE (Squeeze-and-Excitation), SA (Spatial Attention), bi-LSTM (bi-directional LSTM), and DAG (Directed Acyclic Graph). Acronyms mentioned earlier are not reiterated. Methods indicated by references followed by * signify their implementation in this work based on the authors' descriptions. Results for other methods were directly extracted from their respective studies. The maximum value for each metric is highlighted in bold.

| Authors | Time-Frequency Representation | | RC (s) | Technique | Train/Test | Results (%) | | | |
|---|---|---|---|---|---|---|---|---|---|
| | Type | Parameters | | | | *Sen* | *Spe* | *Sco* | *Acc* |
| [22] | STFT | 30 ms | – | HMM | 60/40 | – | – | 39.6 | – |
| [69] | STFT | 500 ms | – | RNN | – (5-fold) | 58.4 | 73.0 | 65.7 | – |
| [121] | STFT | 512 ms | – | HMM SVM | 60/40 | 20.81 | 78.5 | 49.65 | 49.43 |
| [72] | Mel | 250 ms | – | RNN | 80/20 | **62.0** | **84.0** | **74.0** | – |
| [74] | STFT, Wavelet | 20 ms, $D_2 - D_7, A_7$ | – | bi-ResNet | – (10-fold) | 31.1 | 69.2 | 50.2 | 52.8 |
| [73] | STFT, Scalogram | 40 ms | – | CNN | 60/40 | 28.0 | 81.0 | 54.0 | – |
| [78] | STFT | 64 – 128 – 524 ms | – | CNN SVM | – (10-fold) | – | – | – | 65.5 |
| [80] | STFT | 20 ms | – | ResNet NL | 60/40 | 41.3 | 63.2 | 52.3 | – |
| [77] | Mel | 60 ms | – | CNN RNN | 80/20 | – | 58.01 | – | – |
| [81] | STFT | 100 ms | 2.5 | ResNet SE SA | 70/30 | 17.8 | 81.3 | 49.6 | – |
| [79] | STFT | – | 5 | CNN | 70/30 | – | – | – | 74.3 |
| [83] | Mel | – | – | CNN | 60/40 | – | – | – | **80.4** |
| [64] | STFT | 40 ms | – | CNN bi-LSTM | – (5-fold) | 63.0 | 83.0 | 73.0 | – |
| [82] | Wavelet | 30 ms | – | DAG HMM | – | – | – | – | 50.1 |
| [88] | Mel | – | 7 | CNN | 60/40 | 40.1 | 72.3 | 56.2 | – |
| [102] * | STFT | 32 ms 64 filters | 6 | CNN | 80/20 (10-fold) | 51.61 | 65.45 | 58.53 | 60.61 |
| | Mel | | | CNN | | 47.83 | 63.33 | 55.58 | 57.56 |
| | STFT + Mel | | | CNN | | 46.97 | 63.97 | 55.47 | 57.33 |
| [92] | STFT, Log-mel | 32 ms, 50 bins | 8 | ResNet | 60/40 | 37.2 | 79.3 | 58.3 | – |
| [107] | Mel-Spec, MFCC, CQT | 1024 ms | – | CNN + ViT | 60/40 | 36.41 | 78.31 | 57.36 | – |
| This work | Cochleogram | 84 ms 64 filters | 6 | CNN (AlexNet) | 80/20 (10-fold) | 45.12 | 59.75 | 52.43 | 54.48 |
| | | | | CNN (ResNet50) | | 41.78 | 57.78 | 49.78 | 52.31 |
| | | | | CNN (VGG16) | | 53.45 | 68.71 | 61.08 | 62.94 |
| | | | | CNN (Baseline) | | 52.71 | 68.84 | 60.78 | 62.93 |
| | | | | ViT | | 56.77 | 71.37 | 64.03 | 67.99 |

## 6. Conclusions and Future Work

This study introduces an innovative approach that combines the cochleogram with the Vision Transformer (ViT) to enhance the detection and classification of anomalous respiratory sounds. As far as we are aware, this is the initial application of this fusion of TF representation and neural network architecture within this scientific context.

In order to identify the most suitable model for the classification of adventitious respiratory sounds, based on factors such as accuracy, computational cost, and generalizability, a comparative analysis of various conventional neural networks architectures commonly used in sound classification, such as AlexNet, VGG16, ResNet50, and BaselineCNN has been performed. This analysis provides a comprehensive understanding of the strengths and weaknesses of each model, TF representation and neural network architecture, and how they compare to ViT in terms of performance. Ultimately, the findings of this study can inform the development of more effective and efficient approaches to sound classification for medical applications. Results demonstrate that the proposed method, based on cochleogram and ViT, provides the best classification performance, being $Acc$ = 85.9% for wheezes and $Acc$ = 75.5% for crackles detection in the two-class classification scenario and $Acc$ = 67.9% in the four-class classification scenario analysing the entire ICBHI dataset. These results are statistically significant compared to the other evaluated neural networks architectures.

Despite the promising results, leveraging the full potential of the ViT method requires addressing several challenges. One challenge is exploring its applicability to other computer vision tasks, like accurately localizing adventitious sound events within an audio segment. Although initial results are encouraging, further experiments are needed to confirm its effectiveness. Additionally, self-supervised pre-training methods have shown some improvement, but there is still a significant performance gap compared to large-scale supervised pre-training. Bridging this gap through further research is crucial. Moreover, scaling up the ViT method by increasing model size, complexity, and training on larger datasets has proven beneficial for other deep learning models. Doing so is expected to yield higher performance enhancements for the ViT architecture.

**Author Contributions:** Conceptualization, L.D.M., S.G.G. and R.C. methodology, L.D.M., D.M.M. and S.G.G.; software, L.D.M., F.D.G.M. and R.C. validation, L.D.M., F.D.G.M., D.M.M. and R.C. formal analysis, L.D.M., F.D.G.M., S.G.G. and R.C. investigation, L.D.M. and S.G.G.; resources, L.D.M., F.D.G.M. and D.M.M.; data curation, L.D.M., F.D.G.M. and D.M.M.; writing—original draft preparation, L.D.M., F.D.G.M., D.M.M., S.G.G. and R.C. writing—review and editing, L.D.M., F.D.G.M., D.M.M., S.G.G. and R.C. visualization, L.D.M., F.D.G.M., D.M.M. and R.C. funding acquisition, S.G.G. All authors have read and agreed to the published version of the manuscript.

**Funding:** This work was supported in part under grant PID2020-119082RB-{C21,C22} funded by MCIN/AEI/10.13039/501100011033, grant P18-RT-1994 funded by the Ministry of Economy, Knowledge and University, Junta de Andalucía, Spain.

**Institutional Review Board Statement:** Not applicable.

**Informed Consent Statement:** Not applicable.

**Data Availability Statement:** Data are contained within the article.

**Conflicts of Interest:** The authors declare no conflict of interest.